# Spin-orbit readout using thin films of topological insulator $Sb_2Te_3$ deposited by industrial magnetron sputtering

S. Teresi[1], N. Sebe[1], T. Frottier[1], J. Patterson[2], A. Kandazoglou[1],
P. Noël[3], P. Sgarro[1], D. Térébénec[2], N. Bernier[2], F. Hippert [4], J.-P. Attané[1], L. Vila[1], P. Noé[2]
and M. Cosset-Chéneau[1]

[1] Université Grenoble Alpes, CEA, CNRS, INP-G, Spintec, F-38054 Grenoble, France
[2] Université Grenoble Alpes, CEA, LETI, F-38000 Grenoble, France
[3] Department of Materials, ETH Zurich, CH-8093 Zurich, Switzerland
[4] Université Grenoble Alpes, CNRS, Grenoble INP, LMGP, F-38000 Grenoble, France

E-mail: m.n.c.g.cosset-cheneau@rug.nl

Driving a spin-logic circuit requires the production of a large output signal by spin-charge interconversion in spin-orbit readout devices. This should be possible by using topological insulators, which are known for their high spin-charge interconversion efficiency. However, high-quality topological insulators have so far only been obtained on a small scale, or with large scale deposition techniques which are not compatible with conventional industrial deposition processes. The nanopatterning and electrical spin injection into these materials has also proven difficult due to their fragile structure and low spin conductance. We present the fabrication of a spin-orbit readout device from the topological insulator $Sb_2Te_3$ deposited by large-scale industrial magnetron sputtering on $SiO_2$. Despite a modification of the $Sb_2Te_3$ layer structural properties during the device nanofabrication, we measured a sizeable output voltage that can be unambiguously ascribed to a spin-charge interconversion process.

## 1. Introduction

The spin-orbit interaction induced by the spin-charge interconversion opens the way to the creation of spin-logic architectures for low power computing[1]. In these architectures, information is stored in a ferroelectric polarization state coupled to the magnetization direction of a ferromagnetic electrode. The magnetic state of the electrode is then read electrically using the spin to charge interconversion in either heavy metals[2], Rashba interfaces[3] or topological insulators[4]. In these mechanisms, the spin current produced by the ferromagnetic electrode is converted into a transverse charge current thanks to the spin orbit coupling. In order to perform spin-logic operations, the spin-orbit readout device[5] must be able to manipulate the ferroelectric polarization state of a neighboring device using its output voltage such that the devices can be cascaded[6]. Although significant progress has been made in minimizing the ferroelectric switching field[7,8], the required voltages are still much higher than those obtained in current spin-orbit readout devices[9,10,11]. Therefore, their optimization becomes a major challenge for the realization of spin-logic circuits such as the Magnetoelectric Spin-Orbit (MESO) device[6].

Several approaches have been explored for output signal optimization of this spin-orbit readout blocks. It has been observed that downscaling the device leads to a large increase of the output signal[10], while interface engineering is also required to improve the spin-injection efficiency[9,12], and to decrease the shunting of the produced charge current[12,13,14]. To date, most studies have focused on heavy metals, with relatively low spin-charge interconversion efficiencies and resistivities[15]. Therefore, the next natural strategies to optimize the signal of the spin-orbit readout device is to look for materials with higher resistivities[10] and higher spin-charge interconversion efficiencies.

Topological insulators are promising materials for the realization of spin-orbit readout devices with high output voltage. These high-resistivity materials[16] are indeed known to exhibit high spin-charge interconversion efficiencies due to the Edelstein-Effect in their topological surface states[17,18,19], and have already demonstrated their interest for spintronics applications in the context of the spin-orbit torque[20]. The use of these materials for spin-logic circuits is however limited by their fabrication at small-scale by MBE[17,18,21] or by mechanical exfoliation[22,23]. In addition, patterning these materials into nanoscale devices is complicated by their sensitivity to conventional nanofabrication processes[24], so most spin-charge interconversion measurements in topological insulators have been made on microscale devices. Finally, the electrical spin injection into these materials is notoriously difficult due to intermixing effects at the interface with metals[16,25,26] and their low spin conductance caused by their semiconducting nature[27].

$Sb_2Te_3$ is one of the first topological insulator discovered[28], and is known to harbor a high spin-charge interconversion efficiency[29]. This system is also used in industry as part of phase-change memory[30,31], which has led to the development of large-scale deposition techniques for this material[32,33], some of which are industrially compatible, such as magnetron sputtering[34]. However, $Sb_2Te_3$ obtained by large-scale deposition technique has never been studied for the realization of a spin-orbit readout device.

In this paper, we demonstrate the fabrication of a spin-orbit readout device based on $Sb_2Te_3$ deposited using industry compatible processes on $SiO_2$ substrates. We show that a sizeable spin-charge interconversion signal can be obtained in this device by engineering the interface between the $Sb_2Te_3$ and the ferromagnetic spin-injection electrode. Finally, we discuss the effect of our nanofabrication processes on the quality of the $Sb_2Te_3$ film. We show that a ferromagnetic spin injection layer deposited on $Sb_2Te_3$ creates a stress on this material. This leads to the appearance under the ferromagnetic electrode of disorder due to stress-induced disorientation of the initially $c$-axis oriented $Sb_2Te_3$ crystallites.

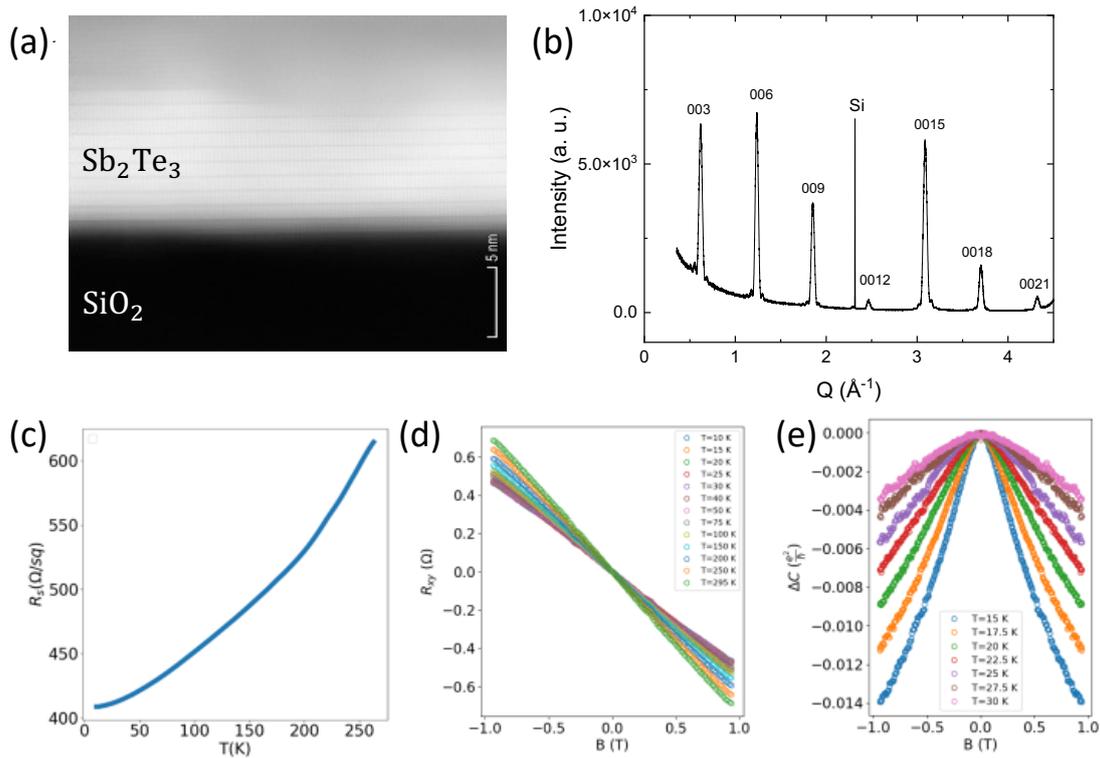

Figure 1: Structural and electrical characterization of the $Sb_2Te_3$ pristine film. (a) Scanning Transmission Electron Microscopy image acquired in High Angle Annular Dark Field (HAADF) mode and (b) XRD pattern measured in the $\theta - 2\theta$ geometry using a Cu ($K_{\alpha 1}$) radiation source ($\lambda = 1.5406$ Å) as a function of $Q = 4\pi \sin(\theta)/\lambda$. The number of ~1 nm thick $Sb_2Te_3$ quintuple layers (QLs) separated by Te-Te van der Waals gaps observed in (a) is lower than that deduced from X-Ray Reflectivity (XRR) because of a partial surface oxidation of the film during Focused Ion Beam (FIB) preparation of the sample lamella prior to the STEM measurement. (c) Sheet resistance, (d) Hall measurement signal and (e) change of conductivity versus magnetic fields measured at different temperatures. The even and odd in field signals were removed from the Hall and weak-antilocalization data, respectively.

## 2. Material characterization

A 15 nm thick $Sb_2Te_3$ film was deposited by magnetron sputtering in an industrial cluster tool at ≈ 250°C on a 300 mm diameter (100) Si wafer, covered with a 100 nm thick thermal $SiO_2$ layer as a bottom insulating layer. Co-sputtering of a stoichiometric $Sb_2Te_3$ target and a Te target was used. As shown previously[31,34], this deposition method compensates for Te desorption and allows to deposit well-oriented $Sb_2Te_3$ films, with Sb and Te planes parallel to the film surface (Figure 1a), due to the formation of a Te atomic layer on top of the 100 nm $SiO_2$ layer. Using X-Ray Reflectivity (XRR), the film thickness is measured to be about 15 nm, while its RMS roughness has been estimated to be 1.4 nm using atomic force microscopy (Figure S1c, supporting information). The $Sb_2Te_3$ film was left uncapped for the purpose of being further integrated in devices, which leads to the formation of a thin oxide layer on the film surface [35].

The structural quality of the film was controlled by X-Ray Diffraction (XRD) patterns acquired in the $\theta - 2\theta$ geometry (Figure 1b). Only *00l* diffraction peaks are detected (hexagonal indexation of the rhombohedral structure of $Sb_2Te_3$). An in-plane diffraction pattern (Figure S1a, supporting information) shows that no preferred in-plane orientation of the crystallites is present. The XRD results show thus that the $Sb_2Te_3$ film is polycrystalline with a fiber texture. The measured hexagonal lattice parameters ($c$ = 3.0537(30) nm and $a$ = 0.42655(10) nm) are consistent with literature values[36]. The structure of $Sb_2Te_3$ can be described as a stacking of quintuple layers (QL) consisting of Te–Sb–Te–Sb–Te planes perpendicular to the [001] direction (hexagonal indexation) and separated by van der Waals-like gaps as visible in Figure 1a. The degree of out-of-plane orientation of the $Sb_2Te_3$ crystallites in the film was determined by analyzing a rocking curve shown in Figure S1b of the supporting information. This curve can be described as the combination of a narrow peak with Full Width at Half Maximum (FWHM) of 0.8°, superimposed on a slightly broader one (FWHM of 3.8°). This indicates the co-existence of well-oriented $Sb_2Te_3$ crystallites, with the *c* axis perpendicular to the film surface, and slightly disoriented crystallites. The detection of Laue oscillations (Figure S2 of the supporting information) indicates the high structural quality of the film. These results are comparable to the ones obtained for MBE-deposited films on amorphous layers[37] or Ge (001) substrates[38], demonstrating the industrial potential of magnetron sputtering for the fabrication of high structural quality topological insulators on a large scale.

We then characterized the transport properties of our $Sb_2Te_3$ films using electrical measurements on as-deposited 15 nm thick films with gold contacts at the corners using the conventional van der Pauw method. The sheet resistance displays a metallic behavior with decreasing temperature (Figure 1c) with a resistivity of $6000 \ \Omega \cdot nm$ at low temperature. This indicates that the $Sb_2Te_3$ bulk is conductive. A predominantly temperature-independent carrier density of $n_s = 10^{20}$ cm$^{-3}$ with a single hole character was extracted from Hall measurements

(Figure 1d), allowing us to extract a mobility of $\mu = 10 \text{ cm}^2 \cdot \text{V}^{-1} \cdot \text{s}^{-1}$ at room temperature. The metallic behavior of $Sb_2Te_3$ has also been observed in MBE-deposited films[39] and bulk crystals[40]. It is attributed to the presence of thermodynamically favored Sb-Te antisite defects, which push the Fermi level into the valence bands of $Sb_2Te_3$[28], thus making its bulk conductive. The carrier density is one order of magnitude higher than that measured in MBE-deposited films of similar thicknesses[29,41], indicating that our magnetron sputtering deposited films present a relatively large density of defects[42]. However, this density is closer to that of MBE films than what has been achieved in widely studied topological insulators such as $Bi_2Se_3$, in which a difference of two orders of magnitude in carrier density between sputtered[43,44] and MBE films[21] was observed. The mobility of our film is low compared to MBE-deposited films, which can be understood by estimating the Ioffe-Regel parameter found to be close to unity, thus indicating an intermediate level of disorder that decreases the mobility[42]. We finally performed weak-antilocalization measurements (Figure 1e). While our low field measurements do not allow us to extract the number of conduction channels, it is still possible to obtain the coherence length[45], which was found to be 50 nm at 10 K by fitting the low field signal using the Hikami-Larkin-Nagaoka formula[46]. This length goes to zero at around 30 K (Figure 2e), as evidenced by the quadratic field dependence of the conductivity variation above this temperature. This dependence of the coherence length with the temperature is consistent with previous weak antilocalization measurements in $Sb_2Te_3$ [47].

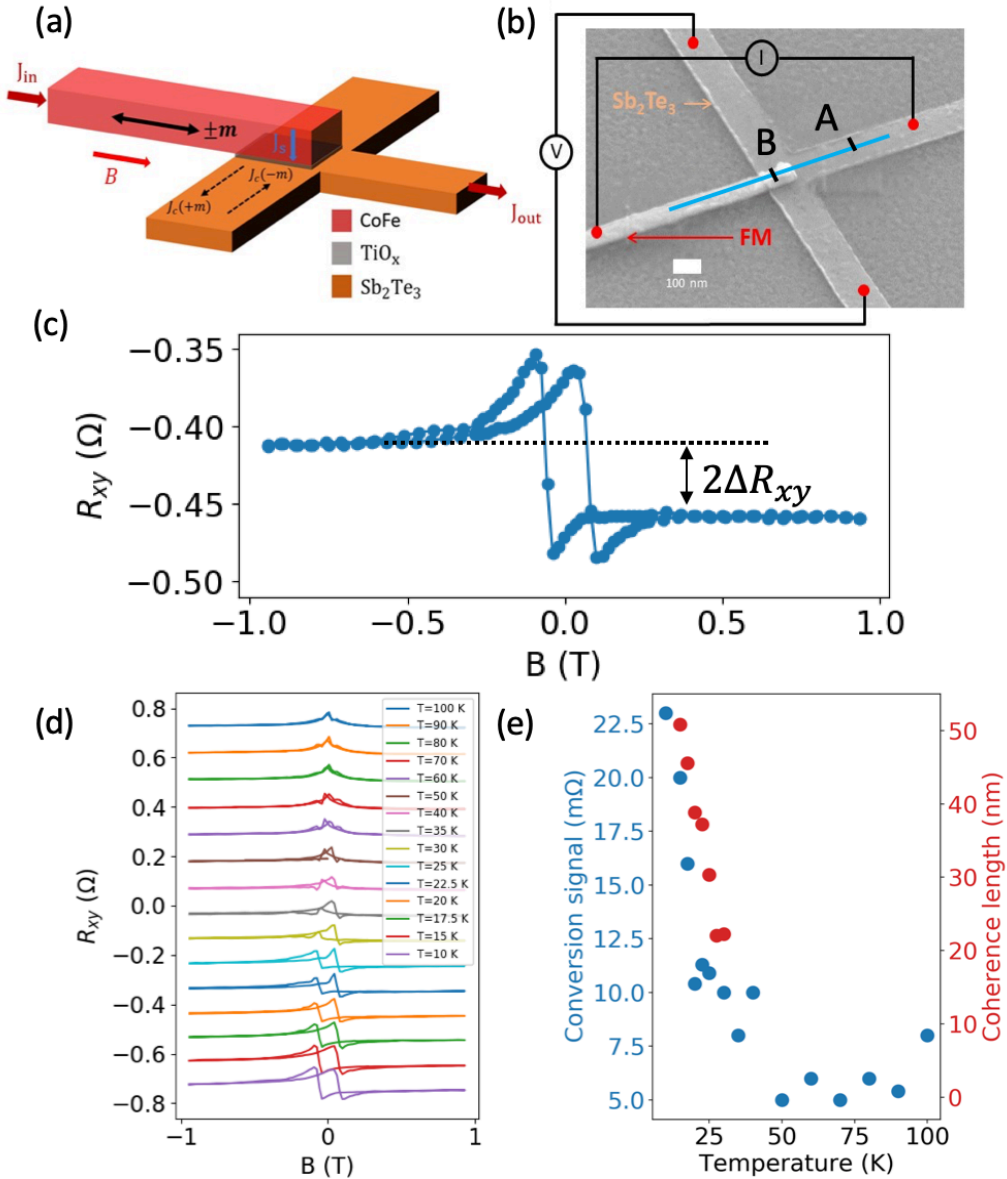

Figure 2: (a) Schematic representation of the spin-orbit readout device. The charge current is injected between the ferromagnetic electrode in red and the inner leg of the T-shaped $Sb_2Te_3$ structure (orange). The polarization of the spin current generated at the interface between the electrode and $Sb_2Te_3$ is along the $\pm m$ magnetization of the electrode (black arrow). The spin to charge interconversion in $Sb_2Te_3$ creates a transverse charge current in the two outer legs of the $Sb_2Te_3$ structure with a sign that depends on the direction of magnetization (black dotted arrows). (b) SEM image of the device presented in (a) with the connections used to flow the charge current and measure the transverse voltage $V_{xy}$. The A and B points corresponds to the Scanning Transmission Electron Microscopy observation area presented in Fig. 3. (c) Transverse resistance versus magnetic field measured at 10 K using the connection configuration presented in (b). (d) Evolution of the transverse resistance versus magnetic field with the temperature. (e) Transverse resistance signal $\Delta R_{xy}$ and coherence length as a function of temperature.

## 3. Spin-orbit readout device fabrication

We then used the $Sb_2Te_3$ film to fabricate the spin-orbit readout device shown in Figure 2a. In this device, a ferromagnetic electrode placed on top of a T-shaped spin-orbit coupling material injects a spin current into the spin-orbit coupling material upon application of an electrical bias current across their interface (Figure 2a). This spin current is then converted into a transverse charge current in the spin-orbit coupling material and detected as a transverse voltage $V_{xy}$ (Figure 2a and Figure 2b). Here, we used $Sb_2Te_3$ as the spin-orbit coupling material, on top of which a 20 nm thick CoFe ferromagnetic electrode is deposited. A 1 nm $TiO_x$ barrier is inserted between the $Sb_2Te_3$ and CoFe layers, as it is known to promote the spin injection in materials with mismatched spin conductance[48]. The T-shape $Sb_2Te_3$ portion is first patterned by conventional Electron-Beam Lithography (EBL). For this purpose, the oxidized top $Sb_2Te_3$ layer is first removed through the EBL resist using a low energy Ion Beam Etching (IBE), and a $TiO_x$(1nm)/CoFe(5nm) hard mask is then deposited on the deoxidized $Sb_2Te_3$ surface by electron beam evaporation, without breaking the vacuum. All metal deposition steps are carried out using electron beam evaporation. The uncovered $Sb_2Te_3$ area is then removed by a soft IBE after lift-off of the hard mask. The ferromagnetic electrode is patterned in a second EBL step. A high-energy IBE is first applied to remove the CoFe oxide layer from the hard mask, before the deposition of the 20 nm thick CoFe electrode, without breaking vacuum. This procedure ensures a good quality of spin contact through $TiO_x$ between CoFe and $Sb_2Te_3$. The hard mask of CoFe on $Sb_2Te_3$ is finally partially removed by IBE, leaving an oxidized thin $TiOx/CoFe$ layer on top of the region of $Sb_2Te_3$ which is not covered by the ferromagnetic electrode. A typical scanning microscopy image of the final device is shown in Figure 2b.

For spin-charge interconversion measurements, we used a standard lock-in amplifier (123 Hz, $I_{bias} = 100$ μA) with the connecting scheme shown in Figure 2b to measure the transverse resistance signal $R_{xy} = V_{xy}/I_{bias}$. The measurements are carried out by applying a magnetic field along the CoFe electrode (Figure 2a), thus allowing to reverse its magnetization direction. A typical transverse signal obtained at 10 K while scanning the magnetic field is shown in Figure 2c. It is constant at high field with a difference of $2\Delta R_{xy} \sim 45$ mΩ between positive and negative field values. The peaks at small field values can be attributed to the planar Hall effect[11], while the baseline is possibly due to a small misalignment of the CoFe electrode with the inner leg of the Sb$_2$Te$_3$ structure[10]. It is important to note that no high-field transverse signal difference was observed in the absence of a TiO$_x$ barrier, evidencing the importance of the barrier to avoid intermixing and/or spin conductivity mismatch. We further observed that $\Delta R_{xy}$ decreases as the temperature increases and vanishes at about 30 K while the low field signal distortions are still present (Figure 2d). $\Delta R_{xy}$ thus follows the temperature dependence of the coherence length measured in unpatterned Sb$_2$Te$_3$ (Figure 2e).

At this point, the attribution of the observed $\Delta R_{xy}$ signal to spin-charge interconversion effects deserves a comment. Indeed, the anomalous Hall effect produced by the charge current flowing vertically into the electrode may produce a signal with similar symmetries[9,49], while the ordinary Hall effect from the stray fields of the ferromagnetic electrode may also produce such a signal[50]. It is possible to separate these different contributions using a combination of systematic geometric dependencies and finite element simulations[16]. Here, these spurious effects can be ruled out using the temperature dependence of $\Delta R_{xy}$ (Figure 2e). As shown in Figure 1d, the Hall effect is almost temperature independent. Moreover, the anomalous Hall effect of CoFe does not vanish at 30 K[51]. Therefore, the ordinary and anomalous Hall effects cannot be at the origin of the observed transverse signal, which therefore comes from a spin-charge interconversion effect in Sb$_2$Te$_3$. The absence of transverse signal when no TiO$_x$ barrier is present proves that a physical separation between the ferromagnetic injection electrode and the Sb$_2$Te$_3$ is necessary to avoid intermixing at the interface[16], as well as spin-backflow and shunting in the ferromagnet[12].

**4. Effect of the patterning on the Sb$_2$Te$_3$ layer structure**

It would be tempting at this point to consider that we succeeded in obtaining an interconversion signal in the topological surface states of a Sb$_2$Te$_3$ film of high structural quality. However, it is known that these materials are sensitive to nanofabrication processes that tend to create disorder. Limiting this disorder is fundamental for electrical transport to be driven by topological surface states[42], and to achieve a high spin-charge interconversion efficiency[17]. To investigate how the nanofabrication process affects the structural quality of Sb$_2$Te$_3$, we performed Scanning Transmission Electron Microscopy (STEM) measurements on our devices. This analysis was performed on two representative regions of the sample (points A and B in Figure 2b).

At point A, Sb$_2$Te$_3$ retains its structural quality as the Sb-Te quintuple layers are clearly visible (Figure 3a). In addition, an Energy-Dispersion X-ray (EDX) spectroscopy map (Figure 3c) shows that the TiO$_x$ barrier is continuous and prevents the Te atoms from diffusing into the CoFe layer, while the Ti, Co and Fe atoms do not diffuse into the Sb$_2$Te$_3$ layer. The EDX maps of all atomic species are shown in Figure S3 of the supporting information. The structure of Sb$_2$Te$_3$ under the CoFe electrode, *i.e.* at point B (Figure 3b) where the interconversion takes place, is, however, more relevant for the spin-charge interconversion. Here, while the Sb and Te planes are still visible, they appear to be blurred and non-parallel to the plane of the substrate. This indicates that the nanofabrication process has induced the disorientation of Sb$_2$Te$_3$ grains whose *c*-axis is now no longer perpendicular to the substrate. In addition, the Energy-dispersive X-ray spectroscopy (EDX) maps of Figure 3d shows that the continuity of the TiO$_x$ layer has been broken in some points and that a large number of magnetic Fe atoms are present in Sb$_2$Te$_3$ (see Figure S3 and Figure S4 in the supporting information for the extensive EDX data set).

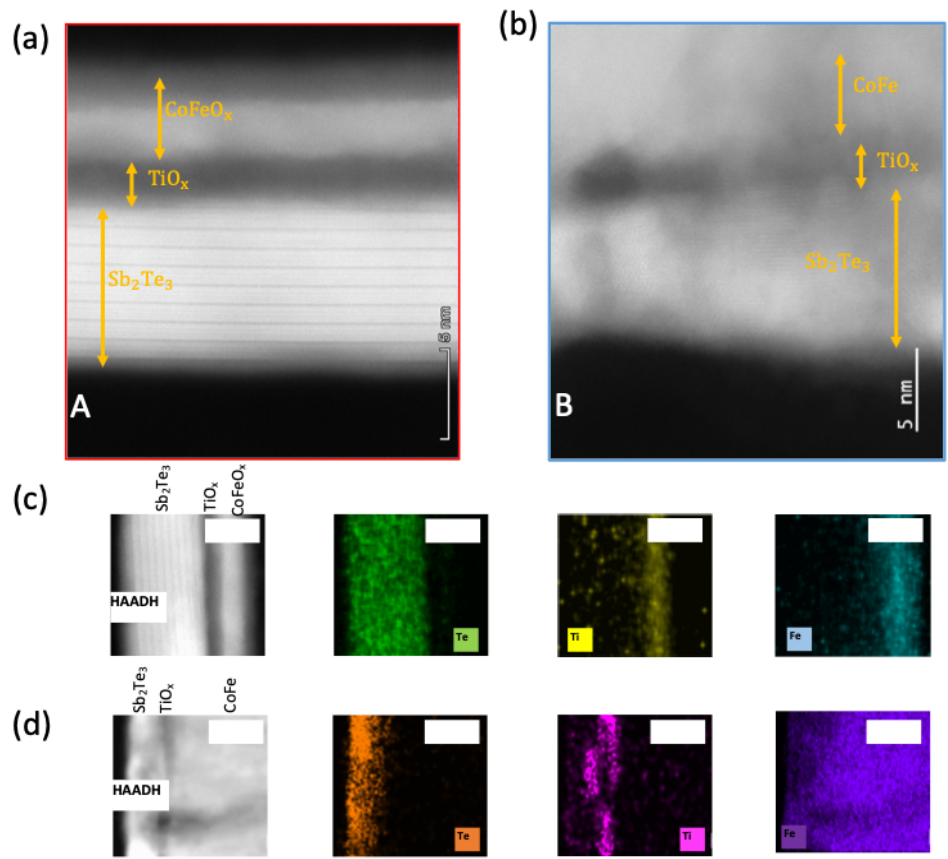

Figure 3: STEM-HAADF images of the FIB lamellae extracted at point A (a) and B (b) in Fig. 2b. Several Sb$_2$Te$_3$ grains are visible in (b) but with no longer preferred orientation with respect to the plane of the substrate. (c) and (d) HAADF images and corresponding EDX maps acquired at point A and B, respectively. The white bars correspond to 7 and 10 nm scales in (c) and (d) respectively.

Points A and B of the device underwent the same EBL processes and have similar in plane feature sizes. Therefore, it is likely that the lower quality of the $Sb_2Te_3$ film at point B compared to point A comes from their different etch/deposit history. In order to investigate the effect of the different etching and deposition steps on the structural quality of $Sb_2Te_3$, we performed XRD measurements in the $\theta - 2\theta$ configuration on unpatterned fullsheet samples that underwent the same etching and deposition steps as points A and B of the devices shown in Figure 2b. For this purpose, we prepared a set of three 1 cm by 1 cm macroscopic samples. Sample 1 is the 15 nm pristine $Sb_2Te_3$ sample that will be used as a reference. Sample 2 is the 15 nm $Sb_2Te_3$ film, on which the low energy IBE step used to remove the $Sb_2Te_3$ surface oxide was applied, followed by the deposition of the $TiO_x$(1 nm)/CoFe(5 nm) layer. Finally, sample 3 underwent the same process as sample 2, with the addition of a high energy etch followed by the deposition of a 20 nm thick CoFe layer. The etching and deposition history of sample 2, respectively 3, is the same as that of point A, respectively B in Figure 2b.

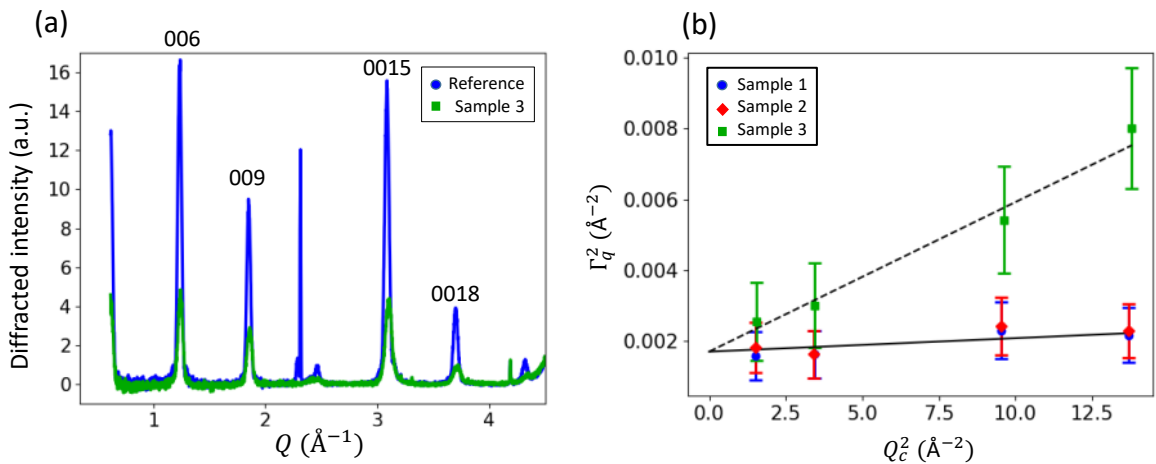

Figure 4: (a) Comparison of the XRD patterns obtained for sample 1 (reference) and for sample 3 in the $\theta-2\theta$ configuration. The measurements were carried out using a Co ($K_\alpha$) radiation source ($\lambda = 1.79$ Å). $Q = 4\pi \sin(\theta)/\lambda$.. (b) Williamson-Hall plot for the reference, sample 2 and sample 3. $\Gamma_Q^2$ is plotted as a function of $Q_c^2$ with $\Gamma_Q = (2\pi/\lambda)\cos(\theta_c)\Gamma$ and $Q_c = (4\pi/\lambda)\sin(\theta_c)$ where $2\theta_c$ is the center of a diffraction peak and $\Gamma$ its FWHM in rds. The linear behavior is fitted using $\Gamma_Q^2 = \epsilon^2 Q_c^2 + (2\pi \times 0.9/D)^2$ with $D$ the diffracting coherent domain of the $Sb_2Te_3$ crystallites $\epsilon$ is related to the random mean square width of the strain distribution. For the reference and sample 2, $\epsilon \sim 5 \times 10^{-3}$, while for sample 3, $\epsilon = 2 \pm 0.5 \times 10^{-2}$.

The measured XRD patterns are shown in Figure 4 and Figure S7a (supporting information). Out-of-plane oriented $Sb_2Te_3$ crystallites are detected in all samples. The rocking curves of samples 1 (reference) and 2 are similar. In sample 3 the rocking curve is a single peak with a FWHM of 1.9°. The diffraction peaks observed in samples 1 and 2 have the same positions and widths. The only change between the two samples is an overall decrease in peak intensity in sample 2, by a factor of 1.35 compared to the reference sample (Figure 7a, supporting information). By contrast, large changes are observed between the XRD patterns of samples 1 and 3. A strong decrease in the intensity of the diffraction peaks is observed in sample 3 compared to the reference sample (Figure 4a). Besides, the diffraction peaks in sample 3 are observed at larger diffraction angles than in the reference and their width is larger, the latter effect being particularly marked for the peaks observed at large diffraction angles. The loss of intensity in sample 3 with respect to sample 1 can be explained by the fact that in sample 3 most of the $Sb_2Te_3$ crystallites are strongly tilted. Only crystallites with the c-axis close to the perpendicular to the substrate are detected in the $\theta-2\theta$ configuration used for acquisition of the XRD patterns. The small reduction of intensity between the reference and sample 2 suggests that tilted crystallites also exist in sample 2 but their number is much smaller than in sample 3. The *c* parameter of the $Sb_2Te_3$ crystallites in sample 3 is slightly smaller (by about 0.4%) than the *c* parameter measured in sample 1 (Figure S7b, supporting information). The *c* parameters of the samples 1 and 2 are identical. This result proves that the 20 nm thick CoFe layer in sample 3 induces a uniform strain on the out-of-plane oriented $Sb_2Te_3$ crystallites, whereas the thinner (5 nm thick) CoFe layer in sample 2 does not induce any significant strain effect.

Besides, A Williamson-Hall analysis of the FWHM of the diffraction peaks [52,53] (Figure 4b) shows that while the size of the $Sb_2Te_3$ crystallites in the direction normal to the substrate remains the same in all samples, within the limits of experimental accuracy, a large inhomogenous strains are present in sample 3 but not in the reference and in sample 2. All these results show that the deposition of a 20 nm thick CoFe layer strongly alters the structural state of the $Sb_2Te_3$ film. It induces a strong tilt of a majority of the $Sb_2Te_3$ crystallites and induces uniform and non-uniform strains on the remaining out of plane-oriented crystallites. We therefore attribute the tilt of the $Sb_2Te_3$ crystallites observed in the HAADF-STEM image at point B in the device (Figure 2b) to the strain applied by the thick CoFe electrode used to inject the spin current into $Sb_2Te_3$. Interestingly, despite the tilt of a majority of the $Sb_2Te_3$ crystallites, a sizeable spin-charge interconversion signal was measured. Assuming that this interconversion is driven only by out-of-plane oriented crystallites, a nanofabrication process which would limit the tilt of $Sb_2Te_3$ crystallites would lead to a much larger transverse signal in the spin-orbit read-out device presented in Figure 3a.

## 5. Conclusion

In conclusion, we have presented the growth of the topological insulator $Sb_2Te_3$ on a large-scale using industrially compatible processes with a structural quality comparable to MBE deposited films. Furthermore, the density of defects comes closer to those of MBE-deposited films than any previous attempts to deposit topological insulators using magnetron sputtering. We then patterned $Sb_2Te_3$ into nanoscale spin-orbit readout devices, with a geometry compatible with the spin-orbit readout block of the MESO device. We obtained a sizeable spin-charge interconversion signal at low temperature by introducing a $TiO_x$ barrier between the CoFe ferromagnetic electrode and the $Sb_2Te_3$ film. Finally, we studied the effect of our nanofabrication processes on the structural quality of the $Sb_2Te_3$ film. We observed that the presence of the thick CoFe layer used as a spin injection electrode creates a stress that induces disorder in the underneath $Sb_2Te_3$ layer.

The recent proposal of a MESO device for low energy spin-logic application has renewed the interest in spin-charge interconversion processes. New types of materials are being investigated to achieve higher spin-charge interconversion efficiencies. However, it is worth noting that while tremendous progresses have been made in this area, a double challenge remains for their use in spin-orbit readout block of the MESO device. First, the materials studied must be grown on a large-scale using industrially compatible processes. Indeed, although interesting from a fundamental point of view, the fabrication of exfoliated and MBE-grown materials cannot be easily scaled up, and large-scale deposition methods must be developed. Second, interconversion studies focus primarily on microscopic[54,55] or macroscopic systems[56,57,58,59]. Although these methods efficiently identify materials of interest from the spin-charge interconversion point of view, their integration in nanodevices with a geometry compatible with the spin-orbit readout block of the MESO device remains scarce[16] and challenging, as illustrated in this paper.

Here, we have addressed both aspects of this problem on a specific material, $Sb_2Te_3$. It is clear that the disorder level in the material, as well as the nanofabrication processes, can still be optimized in order to obtain a larger spin to charge interconversion signal using topological surface state. In addition, these considerations are also relevant for alternative logic architectures such as the recently introduced Ferroelectric Spin-Orbit devices[60] using the ferroelectric polarization of bulk Rashba semiconductors such as GeTe to store the information[61]. This study provides unique insights to overcome the challenges limiting the integration of the recently discovered spin-charge interconversion materials into spin-logic circuits and spin-orbit torques-based memories.

**Supporting Information**

Supporting Information is available from the Wiley Online Library or from the author

**Acknowledgements**

We acknowledge support from the Institut Universitaire de France, from the Project CONTRABASS under Grant Agreements No. ANR-20-CE24-0023 from the Agence Nationale de la Recherche, and form the SPEAR ITN. This project also received funding from the European Union's Horizon 2020 research and innovation program under the Marie Skłodowska-Curie Grant Agreement No. 955671. P.N. acknowledges the support of the ETH Zurich Postdoctoral Fellowship Program 19-2 FEL-61. This project has been partially supported by the European Union's Horizon 2020 research and innovation program under grant agreement No. 824957. The French research group named GdR CHALCO and supported by the CNRS is also thanked for fostering fruitful interactions in the French chalcogenide community. We thank Théo Monniez for the FIB lamalla preparation.

# Supporting information

## Spin-orbit readout using thin films of topological insulator Sb₂Te₃ deposited by industrial magnetron sputtering

S. Teresi, N. Sebe, T. Frottier, J. Patterson, A. Kandazoglou, P. Noël, P. Sgarro, D. Terebenec, N. Bernier, F. Hippert, J.-P. Attané, L. Vila, P. Noé and M. Cosset-Chéneau*

1. Analysis of the pristine film

The XRD analysis on the pristine Sb₂Te₃ film was carried out in the out-of-plane $\theta - 2\theta$ configuration using a Bruker D8 diffractometer equipped with a Ge monochromator selecting the Cu ($K_{\alpha 1}$) radiation ($\lambda = 1.5406$ Å). The degree out-of-plane orientation of the crystallites was determined by measuring a rocking curve around the 009 peak (Figure S1b). The detection of Laue oscillations, shown in Figure S2 for the 006 peak, indicates the high structural quality of the film The thickness of the film was evaluated by analyzing the intensity profile of the 006 peak using

$$I(x) = I_0 \frac{\sin^2(Nx)}{\sin^2(x)} \qquad (1)$$

with $x = \pi(Q - \frac{6}{c})c$ , $Q = (2/\lambda)\sin(\theta)$ and $N$ the number of unit cells in the direction perpendicular to the film. The obtained thickness ($Nc \approx 14$ nm) is close to the value deduced from XRR.

The in-plane XRD pattern shown in Figure S1a was measured using a Rigaku Smartlab diffractometer with the Cu ($K_\alpha$) radiation with no rotation of the film during the measurement. It shows that the film has a fiber texture.

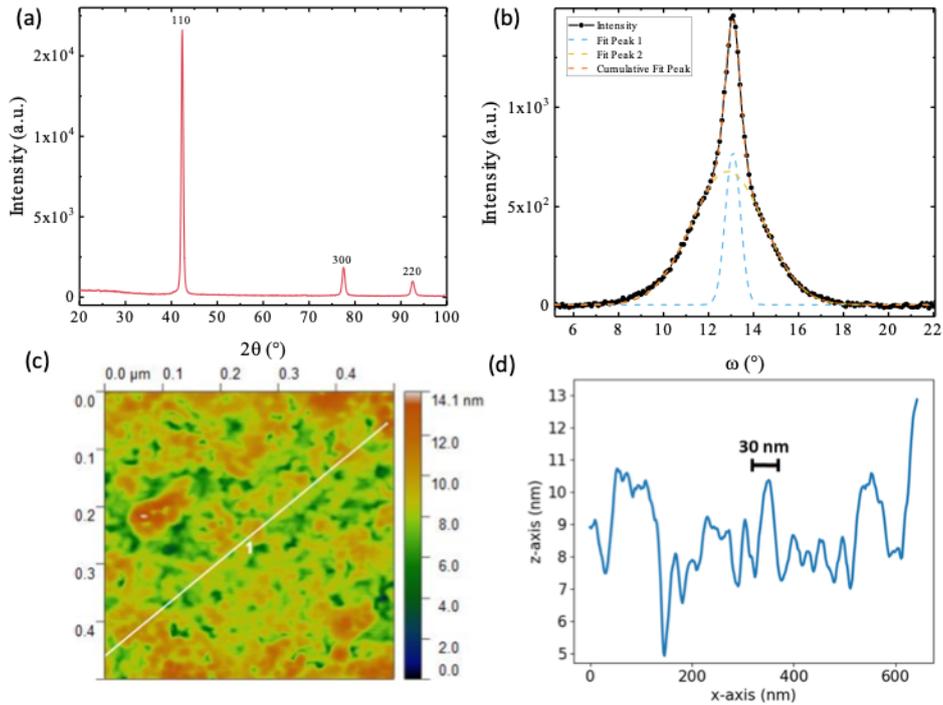

Figure S1: (a) In-plane X-ray diffraction pattern measured on the pristine $Sb_2Te_3$ film using a Cu ($K_α$) radiation source. Only h k 0 diffraction peaks of $Sb_2Te_3$ are detected (hexagonal indexation of the rhombohedral structure of $Sb_2Te_3$). (b) Rocking curve ($ω$ scan at $2θ$ fixed, with ω the incidence angle) measured for the 009 reflection of the pristine $Sb_2Te_3$ film. Dashed lines represent the best fit to the data using two Gaussian curves. The instrument contribution to the FWHM is 0.1°. (c) Atomic Force Microscopy image of the pristine $Sb_2Te_3$ pristine film. (d) Height profile of the $Sb_2Te_3$ film measured along the white line in (c).

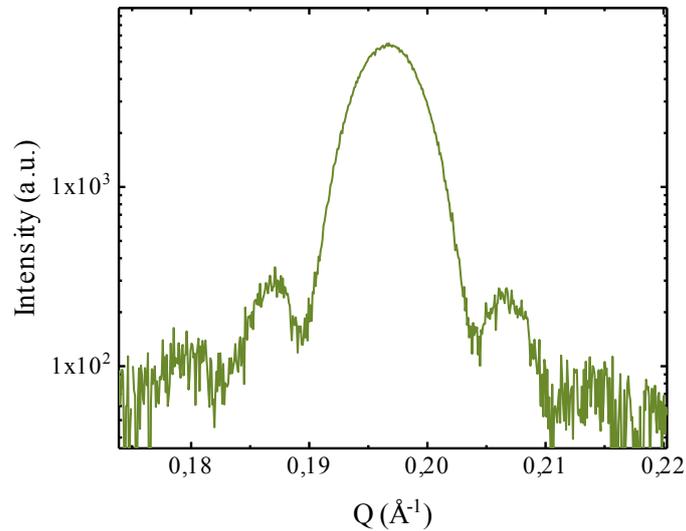

Figure S2: Zoom on the (006) reflection of the out-of-plane x-ray diffraction pattern of the pristine $Sb_2Te_3$ film shown in Fig.1b of the main text. The intensity is plotted as a function of $Q = 2/λ \sin(θ)$. The instrument contribution to the broadening of the peaks is negligible.

## 2. STEM measurements on the devices

The sample was prepared by Ga+ Focused Ion Beam (FIB) milling using a FEI Strata 400 machine. Prior to milling, a thin layer of marker pen is written onto the specimen surface over the region of interest and then a protective W layer is deposited in the FIB machine using ion beam assisted deposition. The specimens are plasma cleaned prior to the TEM analysis in order to remove the ink and leave a region of vacuum near the region of interest such that the protective layers do not interfere with the top layer. For the thin lamella preparation, we used a 16 kV operation voltage for the initial thinning and finished with a low beam energy in the range 5-8 kV to reduce FIB-induced damage. The thin foil was observed in STEM mode at 200 kV using a convergence semi angle of ~ 18 mrad for the incident electron probe in a probe-corrected ThermoFisher Titan Themis microscope equipped with the Super-X detector system for Energy Dispersive X-ray (EDX) spectrometry. The Super-X system comprises four 30mm² windowless silicon drift detectors placed at an elevation angle of 18° from the horizontal with a symmetrical distribution along the beam axis and a 0.64 ± 0.06 sr total solid angle. EDX hypermaps were acquired with a pixel size less than 0.1 nm, a pixel dwell time of ~ 50 µs and for a total acquisition time of ~ 15 min. Hypermaps were processed in the Bruker Esprit v2.2 software using standard TEM recipes for background subtraction and Gaussian peak deconvolution. STEM HAADF images were acquired with a camera length of 86 mm corresponds to inner and outer collection angles of the HAADF detector (Fischione Model M3000) of ~ 78 and 230 mrad. Gun lens and spot size values were selected to provide a probe current of approximately 30 pA. We paid attention to verify that there was no electron beam induced damage by comparing STEM images before and after each STEM/EDX acquisition.

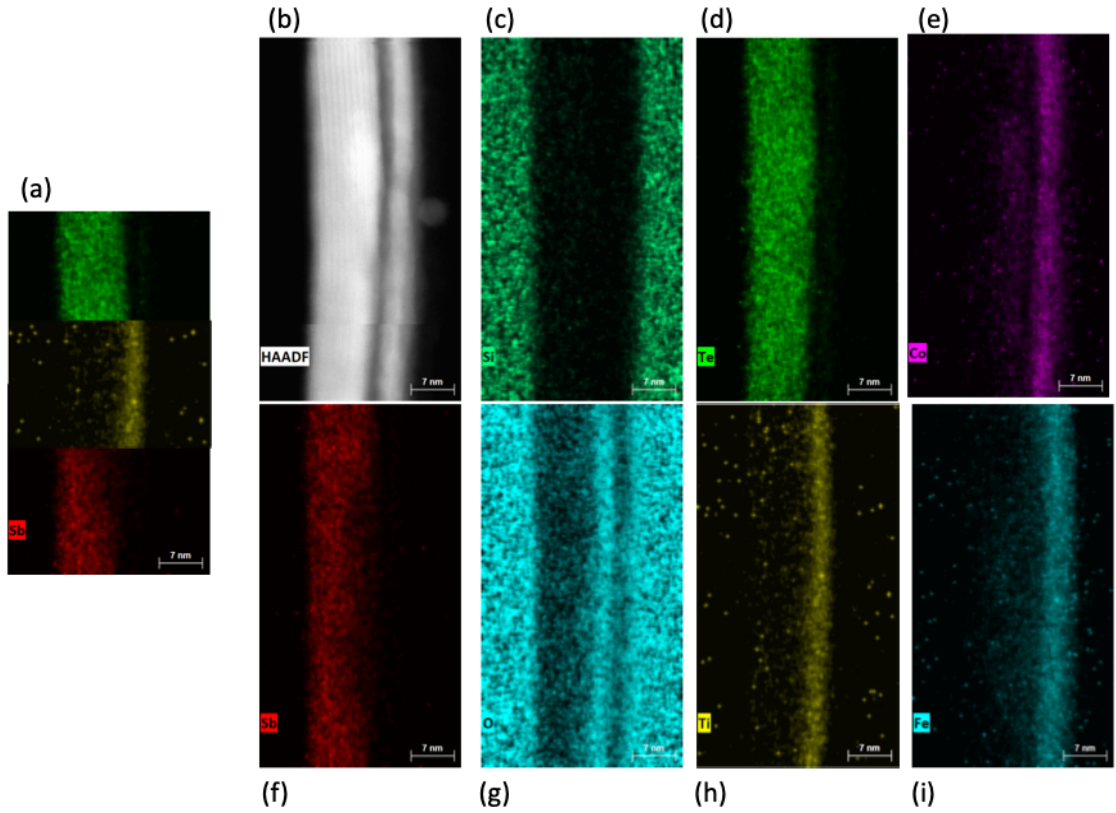

Figure S1: HAADF image and EDX maps of the atomic species present in the device at point A of Fig. 2b. The limited interdiffusion of the Sb (red), Te (green) and Ti (yellow) atoms is visible in (a). The good crystallinity of the $Sb_2Te_3$ can be observed in (b). (c) shows Si on top of the device (right), which is due to the recipe used for the FIB lamella preparation. A small quantity of Co and Fe atoms is present in the $Sb_3Te_2$ layer (e) and (i), which can also be due to the sample lamella preparation. The Ti and CoFe layers are shown to be oxidized in (g), even though the ferromagnetic layers seem to have undergone only a partial oxidation, with the presence of non-oxidized grains as indicated by the darker line in (g) at the position of the CoFe layer.

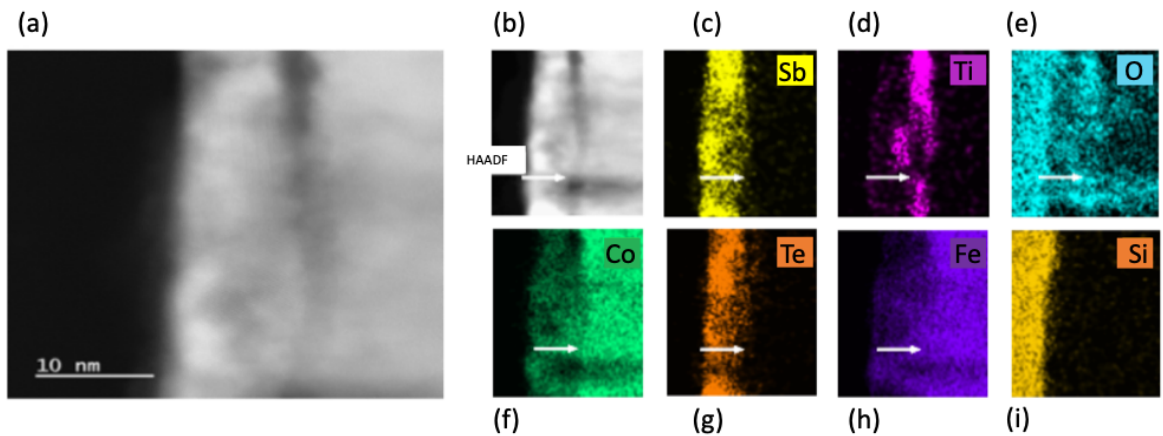

Figure S2: HAADF and EDX map of the atomic species present in the device at point A of Figure 2b in the main text. (a) Shows the presence of several $Sb_2Te_3$ grains under the thick CoFe electrode with difference out of plane orientations. In (d), the TiOx barrier appears broken in the region indicated by the white arrow. This leads to an increased diffusion of Co (f) and Fe (h) atoms in the $Sb_2Te_3$ layer and to a local oxidation of the film.

3. XRD analysis of the patterned Sb$_2$Te$_3$ films

The measurements presented in Figure 4 of the main text where performed in the out-of-plane $\theta - 2\theta$ configuration using a Panalytical Empyrean diffractometer equipped with a cobalt source ($K_\alpha$) radiation ($\lambda = 1.79$ Å) and a $K_\beta$ filter on the diffracted beam. The instrument contribution to the width of the diffraction peaks is negligible. The XRD patterns for samples 1 and 2 are shown in Figure S7a. In Figure S5b we show the Sb$_2$Te$_3$ c-parameter extracted from the fit of the XRD patterns of Figure S5a and Figure 4a of the main text

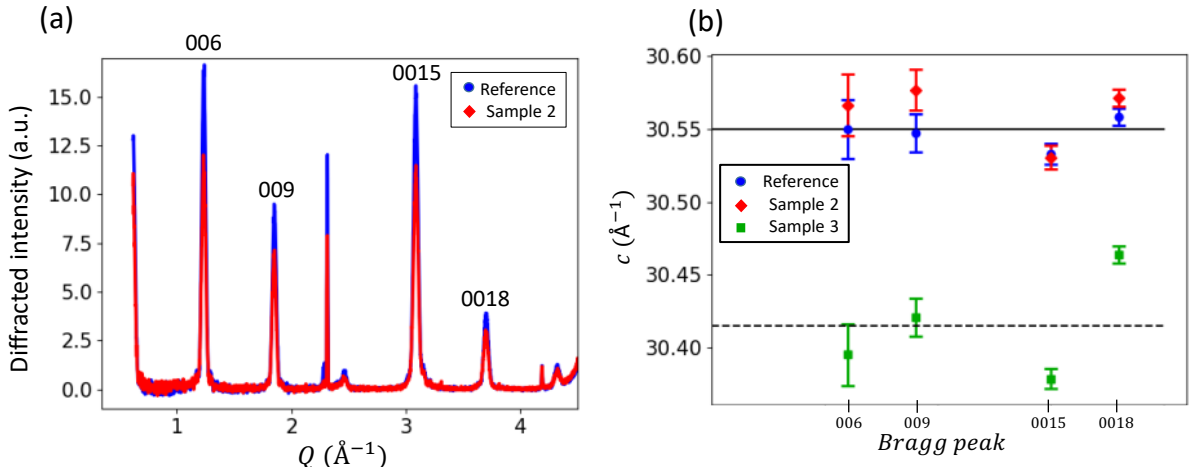

Figure S3: (a) Comparison of the XRD patterns obtained for sample 1 (reference) and for sample 2 in the $\theta-2\theta$ configuration. (b) c-parameters extracted from the XRD measurements presented in Figure 4a of the main text and Figure S7a.

Broadening of the (00l) diffraction peaks as a function of the scattering angle $2\theta$ has been studied for three samples. The peak profiles are close to a Gaussian line and the FWHM of the peak has been obtained by a Gaussian fit, considering the $K_{\alpha 1}$ and $K_{\alpha 2}$ contributions. The instrument contribution to the measured FWHM is negligible. The FWHM in samples 1 and 2 are identical within the error bars and they only slightly increase with $2\theta$. In sample 3 the FWHM steadily increase with $2\theta$. Qualitatively, this trend reveals the existence of a distribution of c values around the mean value determined above. In order to separate the non-uniform strain contribution from the size contribution, we have plotted $\Gamma_Q^2$ as a function of $Q_c^2$ where $\Gamma_Q = (2\pi/\lambda)\cos(\theta_c)\Gamma$ and $Q_c = (4\pi/\lambda)\sin(\theta_c)$ with $2\theta_c$ the center of a diffraction peak and $\Gamma$ its FWHM in rds (Figure 4b in the main text, Williamson-Hall type plot assuming Gaussian profiles). For all films, the data can be fitted by a linear law that can be interpreted as $\Gamma_Q^2 = \epsilon^2 Q_c^2 + (2\pi \times 0.9/D)^2$ where D is the size of a coherently diffracting domain in the direction normal to the film and ε is related to the random mean square width of the strain distribution.

The D value is found the same in all films within the error bars and equal to the thickness (14 nm) of the pristine reference film determined by XRR. For the reference and sample 2, $\epsilon \sim 5 \times 10^{-3}$, while for sample 3, $\epsilon = 2 \pm 0.5 \times 10^{-2}$. Even though the large error bars and small slope prevents a correct estimation of the strain in the reference and sample 2, it is clear from Figure 4b that the strain is much larger in sample 3 than in sample 1 and 2.

**Data Availability Statement**

The data that support the findings of this study are available at
https://doi.org/10.57745/HZDPTT